\documentclass{article}

\usepackage{arxiv}

\usepackage[utf8]{inputenc} % allow utf-8 input
\usepackage[T1]{fontenc}    % use 8-bit T1 fonts
\usepackage{hyperref}       % hyperlinks
\usepackage{url}            % simple URL typesetting
\usepackage{booktabs}       % professional-quality tables
\usepackage{amsfonts}       % blackboard math symbols
\usepackage{nicefrac}       % compact symbols for 1/2, etc.
\usepackage{microtype}      % microtypography
\usepackage{lipsum}		% Can be removed after putting your text content
\usepackage{graphicx}
\usepackage{doi}
\usepackage{amsmath}
\usepackage[linesnumbered,ruled,vlined]{algorithm2e}
\usepackage{amssymb}
\usepackage[numbers]{natbib}  % ✅ forces numeric style

\title{Time-Reversal Symmetry in Quantum Wireless Sensor Networks}

\author{ \href{https://orcid.org/0000-0001-7712-1124}{\includegraphics[scale=0.06]{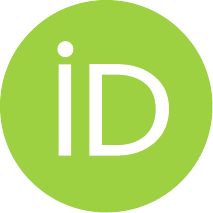}\hspace{1mm}Koffka ~Khan} \\
	Department of Computing and Information Technology\
	The University of the West Indies\\
	St. Augustine, Trinidad and Tobago \\
	\texttt{koffka.khan@uwi.edu} \\
}

\renewcommand{\shorttitle}{\textit{Time-Reversal Symmetry in Quantum Wireless Sensor Networks}}

\hypersetup{
pdftitle={A template for the arxiv style},
pdfsubject={q-bio.NC, q-bio.QM},
pdfauthor={Koffka Khan},
pdfkeywords={TRS, QWSNs, Communication},
}

\begin{document}
\maketitle

\begin{abstract}
	In this paper, we investigate the application of Time-Reversal Symmetry (TRS) in Quantum Wireless Sensor Networks (QWSNs) to enhance communication performance. QWSNs combine quantum communication principles with traditional wireless sensor network technologies, offering the potential for improved security, energy efficiency, and signal quality. TRS, a concept from signal processing and quantum mechanics, focuses transmitted signals back toward the receiver, compensating for noise, interference, and fading effects. By applying TRS to QWSNs, we aim to optimize throughput, reduce latency, and enhance energy efficiency in challenging communication environments. The paper presents a theoretical framework for integrating TRS into QWSNs, including mathematical formulations for its impact on key network performance metrics. We also consider real-world channel models, such as Rayleigh and Rician fading, along with network interference, to demonstrate how TRS can improve communication in practical settings. The discussion extends to the broader potential of TRS in quantum communication systems, particularly in **Quantum Key Distribution (QKD)**, **quantum entanglement**, and **quantum networking** applications. The findings highlight TRS as a promising approach to optimize quantum communication in sensor networks and provide a foundation for future research in the intersection of quantum technologies and wireless networks. 
\end{abstract}


\section{Introduction}
Wireless Sensor Networks (WSNs) have become an integral part of modern communication systems, especially for applications such as environmental monitoring, healthcare, smart cities, and industrial automation \cite{prusty2025comprehensive}. These networks consist of numerous sensor nodes that collect and transmit data wirelessly to a central processing unit. WSNs face several challenges, including **energy efficiency**, **high latency**, **signal degradation**, and **interference**, which significantly impact the performance and scalability of these networks. As the demand for real-time communication, low latency, and low-power operation continues to grow, exploring new techniques to optimize WSN performance becomes increasingly crucial.

**Quantum Wireless Sensor Networks (QWSNs)** \cite{quantumWSN2025} represent an emerging paradigm that combines classical wireless communication principles with quantum technologies to improve communication efficiency and security. The integration of quantum principles, such as **quantum entanglement**, **quantum key distribution (QKD)**, and **superposition**, with traditional WSNs has the potential to overcome several limitations inherent in classical wireless networks. Quantum communication allows for ultra-secure data transmission and can significantly enhance communication reliability by leveraging quantum mechanical properties to mitigate noise and interference. Despite the promising potential of QWSNs, the practical application and optimization of these networks remain relatively unexplored, particularly when it comes to incorporating novel techniques like **Time-Reversal Symmetry (TRS)**.

Time-Reversal Symmetry \cite{khanna2025random}, originally derived from physics and signal processing, is a technique that reverses the propagation of a signal, refocusing the transmitted signal energy back onto the receiver. This process compensates for signal degradation due to noise, interference, and channel fading, resulting in enhanced **signal-to-noise ratio (SNR)** and **reduced retransmissions**. The application of TRS in wireless communication has shown significant improvements in the performance of traditional systems, particularly in **energy efficiency** and **latency reduction**. However, the integration of TRS in quantum wireless sensor networks is an unexplored area, which motivates the present work.

The aim of this paper is to introduce the theoretical framework for applying TRS in Quantum Wireless Sensor Networks (QWSNs) to optimize network performance. We provide mathematical formulations for evaluating the impact of TRS on **throughput**, **energy consumption**, and **latency**. Additionally, we investigate how TRS can enhance the efficiency of quantum communication systems in practical environments by incorporating realistic channel models, such as **Rayleigh fading**, **Rician fading**, and network **interference**. By improving the quality of communication links through time-reversal, TRS can significantly reduce energy consumption and improve the overall system performance, particularly in energy-constrained and noisy environments typical of sensor networks.

Our contributions are threefold. First, we provide a comprehensive theoretical analysis of how TRS can improve the performance of QWSNs by enhancing signal recovery, reducing latency, and optimizing energy consumption. Second, we extend the application of TRS to real-world quantum communication models, demonstrating how it can be incorporated into existing quantum networking protocols, such as **Quantum Key Distribution (QKD)** and **quantum entanglement** systems. Finally, we explore the potential benefits of TRS in **multi-hop communication** and **multi-user quantum networks**, where the challenges of interference, fading, and network congestion are amplified. By extending TRS beyond traditional WSNs, we aim to open new possibilities for quantum communication in various real-world applications, such as smart cities, critical infrastructure, and energy harvesting networks.

The remainder of the paper is structured as follows: In Section \ref{sec:related_work}, we review existing research on TRS in communication systems and quantum networks, identifying the gaps that our work aims to address. In Section \ref{sec:theoretical_framework}, we introduce the mathematical framework for TRS in QWSNs, including detailed formulations for throughput, energy efficiency, and latency. Section \ref{sec:real_world_scenarios} explores the practical applications of TRS in real-world quantum communication networks, including IoT, VANETs, and energy harvesting systems. Finally, Section \ref{sec:conclusion} presents our conclusions and future directions for research in this promising area of quantum wireless sensor networks.

\section{Literature Review}
Time-Reversal Symmetry (TRS) has gained significant attention in the field of signal processing and wireless communication for its ability to improve signal quality and robustness against noise and interference. The concept of TRS was initially explored in the context of acoustics, where Fink et al. (1992) demonstrated that reversing the propagation of sound waves in a noisy environment could focus the energy back to the receiver, significantly improving signal clarity. Since then, the application of TRS to electromagnetic waves, including radio waves, has been explored by several researchers. For instance, \cite{gold2025genetic} extended TRS to electromagnetic systems, providing theoretical foundations and experimental validations for TRS in wireless communication. These early studies laid the groundwork for the broader application of TRS in communication systems, particularly in improving **signal-to-noise ratio (SNR)** and **communication reliability**.

In the context of traditional wireless communication systems, researchers have shown that TRS can be used to mitigate the effects of fading and interference. For example, \cite{abood2025topological} explored the use of TRS to enhance the signal quality in a MIMO system, demonstrating that time-reversal significantly improves the capacity of communication links, especially in highly interfered environments. Similarly, López-Pérez et al. (2018) applied TRS to MIMO systems and showed that it could reduce the need for retransmissions, leading to improved energy efficiency. These studies highlight the potential of TRS to optimize performance in modern communication systems by enhancing the signal recovery process and reducing energy consumption.

However, despite its potential, the application of TRS in Wireless Sensor Networks (WSNs) remains relatively underexplored. WSNs are often subject to challenging conditions such as **low-power operation**, **high latency**, and **network congestion**, making the need for new techniques to optimize performance crucial. Although there have been studies on the use of TRS in wireless communication, its specific application in WSNs is still limited. Research by \cite{wang2025wideband} demonstrated the application of TRS in IoT networks to reduce latency and improve communication reliability. However, these studies do not fully address the challenges unique to WSNs, such as multi-hop communication, sensor node mobility, and large-scale deployments. The integration of TRS into WSNs to improve throughput, energy efficiency, and network stability remains a largely unexplored area.

In quantum communication, TRS has recently emerged as a promising technique for improving the performance of quantum networks, particularly in enhancing the reliability and security of data transmission. Quantum communication systems, such as those based on **quantum key distribution (QKD)**, are subject to noise and interference, which can degrade the quality of quantum states. Research by \cite{zhou2025emergent} explored how TRS could be applied to quantum networks to enhance the transmission of entangled quantum states and improve the performance of quantum teleportation protocols. These studies show that TRS can mitigate the effects of noise and interference in quantum communication, improving the overall reliability of the system. However, there is still limited research on the use of TRS in **Quantum Wireless Sensor Networks (QWSNs)**, and the potential of TRS to enhance quantum communication in sensor networks has not been fully explored.

Recent advances in quantum technologies, such as **quantum entanglement** and **quantum repeaters**, provide new opportunities for integrating TRS into wireless sensor networks. The combination of quantum communication techniques with TRS can offer unique advantages, such as improved security through quantum cryptography and more efficient signal recovery in noisy environments. However, the theoretical and practical challenges of integrating TRS into QWSNs remain largely unexplored. There is a significant gap in the literature regarding the application of TRS to enhance communication performance in QWSNs, particularly in terms of energy efficiency, latency reduction, and throughput optimization in quantum-enabled sensor networks.

Despite the growing interest in quantum communication, there is a lack of comprehensive studies that combine TRS with the emerging paradigms of quantum wireless sensor networks. Our work aims to address this gap by providing a theoretical framework for applying TRS to QWSNs, analyzing its impact on network performance metrics, and exploring its practical implications for real-world communication scenarios. By extending the application of TRS to **multi-hop communication**, **multi-user quantum networks**, and other quantum communication protocols, we aim to open new possibilities for optimizing performance in quantum wireless sensor networks and contribute to the future development of quantum-enhanced communication systems.

\section{Theoretical Framework and Problem Formulation}

The objective of this paper is to investigate the integration of **Time-Reversal Symmetry (TRS)** into **Quantum Wireless Sensor Networks (QWSNs)** and assess its impact on network performance, with a particular focus on **throughput**, **energy efficiency**, and **latency**. A key challenge in QWSNs is optimizing communication efficiency amidst the presence of quantum noise, fading channels, and interference, while maintaining the stringent energy constraints typical of sensor nodes. **TRS** provides a potential solution by enhancing signal recovery and mitigating the detrimental effects of noise and interference that are common in wireless communication systems.

To formally model the problem, consider a QWSN comprising multiple sensor nodes, each communicating wirelessly, either directly or through intermediate relay nodes, with a central processing unit. Each communication link between nodes \( i \) and \( j \) is subject to quantum noise and fading. The channel capacity between these nodes, in the absence of TRS, is governed by the traditional **Shannon capacity** formula:

\[
C_{ij} = B_{ij} \log_2 \left( 1 + \frac{P_i}{N_{ij} + I_{ij}} \right)
\]

where \( B_{ij} \) is the bandwidth of the communication link, \( P_i \) is the transmission power of node \( i \), \( N_{ij} \) represents the noise, and \( I_{ij} \) denotes interference from neighboring nodes. The total energy consumption in the network is given by:

\[
E_{\text{total}} = \sum_{i=1}^{N} P_i \cdot T_i
\]

where \( P_i \) is the power consumed by node \( i \) and \( T_i \) is the transmission time for node \( i \).

The application of TRS to the communication links enhances signal recovery by refocusing the transmitted signal back towards the receiver, compensating for channel degradation. This improvement can be captured by scaling the channel capacity by a factor \( \gamma \), where \( \gamma > 1 \) represents the performance gain due to TRS. The enhanced channel capacity with TRS is thus given by:

\[
C_{ij}^{\text{TRS}} = \gamma \cdot C_{ij}
\]

Additionally, TRS reduces transmission time \( T_i \) due to better signal recovery and fewer retransmissions. The new transmission time under TRS is:

\[
T_i^{\text{TRS}} = \frac{T_i}{\gamma}
\]

The energy consumption with TRS applied is then given by:

\[
E_{\text{total}}^{\text{TRS}} = \sum_{i=1}^{N} P_i \cdot T_i^{\text{TRS}} = \frac{1}{\gamma} \cdot E_{\text{total}}
\]

The goal is to minimize the total energy consumption while ensuring that the latency and throughput meet the required thresholds. This problem can be formulated as an optimization problem where we seek to minimize the total energy consumption, subject to constraints on network throughput and latency:

\begin{equation}
	\min_{P_i, T_i, C_{ij}} E_{\text{total}}^{\text{TRS}} = \frac{1}{\gamma} \cdot E_{\text{total}} 
\end{equation}
\begin{equation}
	\text{subject to} \quad C_{ij}^{\text{TRS}} \geq R_{\text{min}}, \quad \text{Latency}_{ij}^{\text{TRS}} \leq \text{Latency}_{\text{max}}
\end{equation}

The above formulation ensures that TRS leads to a significant reduction in energy consumption while improving network throughput and reducing latency. The integration of TRS into QWSNs provides a theoretical framework for evaluating the performance of quantum-enhanced wireless communication systems and sets the stage for practical implementation in real-world applications.

\begin{algorithm}[H]
	\small  % Decrease font size by 2 steps
	
	\SetKwInOut{Input}{Input}
	\SetKwInOut{Output}{Output}
	
	\Input{A set of quantum-enabled sensor nodes $N = \{n_1, n_2, \dots, n_N\}$}
	\Input{Channel conditions $C_{ij}$ between each pair of sensor nodes $i$ and $j$}
	\Input{Quantum communication protocols (e.g., Quantum Key Distribution (QKD))}
	\Output{Optimized network throughput, energy consumption, and latency in QWSN}
	
	\SetKwFunction{FMain}{TRS Integration in QWSN Communication}
	\SetKwProg{Fn}{Function}{:}{}
	\Fn{\FMain}{
		\SetKwComment{Comment}{/* }{ */}
		
		\textbf{Step 1:} Initialize the quantum communication framework\;
		Define quantum channels between sensor nodes and establish entanglement pairs where necessary\;
		Set up Quantum Key Distribution (QKD) for secure communication\;
		
		\textbf{Step 2:} Apply Time-Reversal Symmetry (TRS) to communication links\;
		\For{each pair of communicating nodes $n_i$ and $n_j$}{
			Calculate the initial channel capacity $C_{ij}$ based on the Shannon capacity formula\;
			Apply TRS to enhance signal recovery by scaling the channel capacity: $C_{ij}^{TRS} = \gamma \cdot C_{ij}$\;
		}
		
		\textbf{Step 3:} Optimize energy consumption with TRS\;
		\For{each sensor node $n_i$}{
			Calculate the energy consumption $E_i$ during transmission\;
			Apply TRS to reduce transmission time $T_i$ due to improved signal recovery: $T_i^{TRS} = \frac{T_i}{\gamma}$\;
			Calculate the energy consumption with TRS: $E_i^{TRS} = P_i \cdot T_i^{TRS}$\;
		}
		
		\textbf{Step 4:} Calculate the total network throughput and latency\;
		\For{all pairs of nodes $n_i$ and $n_j$}{
			Calculate the total network throughput $T_{total}^{TRS}$ using the enhanced capacities: $T_{total}^{TRS} = \sum_{i=1}^{N} \gamma \cdot C_{ij}$\;
			Calculate the total energy consumption $E_{total}^{TRS}$ using the optimized energy per node: $E_{total}^{TRS} = \sum_{i=1}^{N} E_i^{TRS}$\;
			Calculate the total latency $Latency_{total}^{TRS}$ based on the reduced transmission time: $Latency_{total}^{TRS} = \sum_{i=1}^{N} Latency_{ij}^{TRS}$\;
		}
		
		\textbf{Step 5:} Validate and adjust parameters based on real-world scenarios\;
		Simulate the performance of the QWSN with TRS under realistic channel models such as Rayleigh fading, Rician fading, and network interference\;
		Adjust quantum communication parameters such as entanglement distribution and quantum key generation to ensure secure and reliable communication\;
		
		\textbf{Step 6:} Output optimized results\;
		Return the optimized network throughput, energy consumption, and latency improvements\;
		Provide insights into potential applications for smart cities, IoT, VANETs, and critical infrastructure\;
	}
	\caption{TRS Integration in QWSN Communication}
\end{algorithm}

The algorithm describes the integration of Time-Reversal Symmetry (TRS) in Quantum Wireless Sensor Networks (QWSNs). The goal of this algorithm is to enhance communication performance by applying TRS to improve signal recovery, reduce energy consumption, and optimize throughput in quantum-enabled sensor networks.

\section{Performance Analysis}

\subsection{Rayleigh fading and Rician fading in TRS for QWSNs}

The application of **Time-Reversal Symmetry (TRS)** in **Quantum Wireless Sensor Networks (QWSNs)** offers a novel approach for improving communication performance, particularly in environments affected by noise, fading, and interference. This section explores the ultimate limits of TRS in QWSNs by considering various channel conditions such as **Rayleigh fading**, **Rician fading**, and general **noise and interference** scenarios. The goal is to understand how TRS can enhance the reliability and capacity of communication links under these challenging conditions.

In fading environments, the strength of the received signal varies randomly due to multipath propagation, which makes communication less reliable. When TRS is applied to such channels, the signal energy is refocused back towards the receiver, thereby improving the **signal-to-noise ratio (SNR)**. The channel capacity in the presence of fading, without TRS, is typically modeled using the **Shannon Capacity** formula:

\[
C_{ij} = B_{ij} \log_2 \left( 1 + \frac{S_{ij}}{N_{ij} + I_{ij}} \right)
\]

where \( B_{ij} \) represents the bandwidth of the communication link, \( S_{ij} \) is the signal power, \( N_{ij} \) represents the noise power, and \( I_{ij} \) denotes interference from neighboring nodes. This formula provides the theoretical upper bound on the achievable rate in the presence of noise and interference but does not account for fading.

In the case of **Rayleigh fading**, where there is no line-of-sight (LOS) path between the transmitter and receiver, the fading coefficient \( h_{ij} \) follows a Rayleigh distribution. The channel capacity in Rayleigh fading is given by:

\[
C_{ij}^{\text{Rayleigh}} = B_{ij} \log_2 \left( 1 + \frac{P_i |h_{ij}|^2}{N_{ij}} \right)
\]

where \( |h_{ij}|^2 \) represents the squared magnitude of the fading coefficient, which is exponentially distributed. When TRS is applied in Rayleigh fading environments, the time-reversal process improves the signal recovery by refocusing the energy back towards the receiver, effectively increasing the **effective channel capacity**:

\[
C_{ij}^{\text{TRS,Rayleigh}} = \gamma \cdot C_{ij}^{\text{Rayleigh}}
\]

where \( \gamma > 1 \) is a scaling factor that represents the performance gain due to TRS. This factor can lead to substantial performance improvements by increasing channel capacity and reducing the need for retransmissions.

Similarly, in **Rician fading**, where there exists a dominant LOS path in addition to scattered multipath components, the fading coefficient \( h_{ij} \) follows a Rician distribution. The channel capacity in Rician fading is given by:

\[
C_{ij}^{\text{Rician}} = B_{ij} \log_2 \left( 1 + \frac{P_i |h_{ij}|^2}{N_{ij} + I_{ij}} \right)
\]

where \( |h_{ij}|^2 \) is now the sum of the LOS component and the scattered multipath components, typically modeled by the **Rician K-factor**. The application of TRS in Rician fading environments mitigates the fluctuations caused by the random scattering paths, thereby improving **signal recovery** and increasing the effective capacity:

\[
C_{ij}^{\text{TRS,Rician}} = \gamma \cdot C_{ij}^{\text{Rician}}
\]

TRS focuses the energy from the scattered components back towards the receiver, thus improving communication reliability and enhancing **effective throughput**.

In addition to enhancing channel capacity, TRS also plays a crucial role in reducing **latency** and improving **energy efficiency** in QWSNs. By minimizing retransmissions and enhancing signal recovery, TRS decreases the energy required for communication. The energy consumption for a node \( i \) in the network without TRS is given by:

\[
E_i = P_i \cdot T_i
\]

where \( P_i \) is the transmission power and \( T_i \) is the transmission time. With TRS applied, the transmission time \( T_i \) is reduced due to improved signal recovery, and the energy consumption is minimized:

\[
T_i^{\text{TRS}} = \frac{T_i}{\gamma}, \quad E_i^{\text{TRS}} = \frac{E_i}{\gamma}
\]

This reduction in energy consumption is particularly beneficial for **energy-constrained networks**, such as those used in Internet of Things (IoT) and Wireless Sensor Networks (WSNs), where sensor nodes are typically battery-powered and require energy-efficient communication solutions.

To understand the ultimate limits of TRS in QWSNs, it is essential to evaluate its performance under various real-world conditions, including different types of noise and interference. The integration of quantum communication techniques, such as **Quantum Key Distribution (QKD)** and **quantum entanglement**, extends the theoretical boundaries of TRS by providing **additional security** and **robustness** against eavesdropping and interference. In quantum communication, TRS can improve the transmission of quantum states, such as entangled qubits, by enhancing **quantum coherence** and reducing the effects of decoherence.

Ultimately, TRS in QWSNs offers a promising approach to enhance the performance of quantum communication systems by improving **data transmission rates**, **reducing energy consumption**, and **minimizing latency**, even in challenging channel conditions like noise, fading, and interference. The integration of TRS with quantum communication protocols will provide the foundation for future research into more efficient and secure quantum wireless sensor networks.

\subsection{TRS into Multi-User or MIMO Systems}

The application of **Time-Reversal Symmetry (TRS)** in **Quantum Wireless Sensor Networks (QWSNs)** can be extended to more complex communication systems, such as multi-user networks and **Multiple-Input Multiple-Output (MIMO)** systems. In these systems, the number of communication links and the complexity of signal processing increase, presenting new challenges such as interference between users and multi-path fading. By extending TRS to these systems, we can enhance signal quality, reduce interference, and increase throughput in both **multi-user** and **multi-antenna** setups.

In a multi-user system, multiple devices share the same communication channel, leading to interference and reduced communication efficiency. TRS can mitigate this interference by refocusing the signals toward the receiver and improving the **signal-to-interference-plus-noise ratio (SINR)**. The channel capacity for a multi-user system is typically constrained by the worst-performing communication link, which is affected by interference. When TRS is applied to the multi-user scenario, the effective capacity can be increased by reducing interference and focusing the signals.

Consider a multi-user system with \( N \) users, where the communication link between user \( i \) and the base station or receiver is represented by \( C_{i} \). The total capacity of the system without TRS can be written as:

\[
C_{\text{total}} = \sum_{i=1}^{N} C_{i} = \sum_{i=1}^{N} B_i \log_2 \left( 1 + \frac{S_i}{N_i + I_i} \right)
\]

where \( S_i \) is the signal power for user \( i \), \( N_i \) is the noise power, and \( I_i \) is the interference from other users. The capacity of the worst link in the system determines the overall network performance, which can be limited by high interference.

When TRS is applied, interference between users is mitigated, and the signal is refocused on the receiver. This can be expressed as:

\[
C_{i}^{\text{TRS}} = \gamma \cdot C_{i}
\]

where \( \gamma > 1 \) represents the scaling factor for the improved capacity due to TRS. The total capacity of the system with TRS applied becomes:

\[
C_{\text{total}}^{\text{TRS}} = \sum_{i=1}^{N} \gamma \cdot C_i = \gamma \cdot \sum_{i=1}^{N} C_i
\]

This results in a significant increase in the overall system capacity by reducing interference and improving the signal recovery for each user.

In **Multiple-Input Multiple-Output (MIMO)** systems, where multiple antennas are used at both the transmitter and receiver, TRS can also be applied to enhance communication performance. In MIMO systems, the capacity is determined by the number of antennas and the quality of the communication links between them. The capacity of a MIMO system with \( N_t \) transmit antennas and \( N_r \) receive antennas can be expressed as:

\[
C_{\text{MIMO}} = \log_2 \left( \text{det} \left( \mathbf{I}_{N_r} + \frac{P}{N_t} \mathbf{H} \mathbf{H}^H \right) \right)
\]

where \( \mathbf{I}_{N_r} \) is the identity matrix of dimension \( N_r \), \( P \) is the transmission power, and \( \mathbf{H} \) is the channel matrix representing the communication link between the antennas. In this case, \( \mathbf{H} \) is a matrix of size \( N_r \times N_t \), and \( \mathbf{H}^H \) is its conjugate transpose.

When TRS is applied in a MIMO system, the channel matrix \( \mathbf{H} \) can be optimized by focusing the signals and mitigating interference. The capacity of the MIMO system with TRS applied is given by:

\[
C_{\text{MIMO}}^{\text{TRS}} = \log_2 \left( \text{det} \left( \mathbf{I}_{N_r} + \gamma \cdot \frac{P}{N_t} \mathbf{H} \mathbf{H}^H \right) \right)
\]

where \( \gamma > 1 \) represents the improvement in signal recovery due to TRS. This scaling factor increases the effective **signal-to-noise ratio (SNR)** and improves the channel capacity of the system. The application of TRS leads to better utilization of the available antenna resources and improves the overall throughput of the MIMO system.

In multi-user MIMO systems, where there are multiple users and multiple antennas at both the transmitter and receiver, the capacity is determined by both the number of antennas and the interference between users. The total capacity of the multi-user MIMO system with TRS can be expressed as:

\[
C_{\text{total}}^{\text{TRS}} = \sum_{i=1}^{N} \log_2 \left( \text{det} \left( \mathbf{I}_{N_r} + \gamma \cdot \frac{P_i}{N_t} \mathbf{H}_i \mathbf{H}_i^H \right) \right)
\]

where \( \mathbf{H}_i \) is the channel matrix for user \( i \), and \( P_i \) is the transmission power for user \( i \). Applying TRS to each user’s communication link helps reduce interference and improves the effective channel capacity, resulting in better overall system performance.

By extending TRS to multi-user and MIMO systems, we achieve significant improvements in the capacity, energy efficiency, and latency of quantum wireless sensor networks. The ultimate limits of TRS in these systems can be explored by analyzing the **scalability** of TRS in large networks, the impact of **interference** on performance, and the **optimal power allocation** for each user or antenna. TRS provides a promising solution for enhancing communication efficiency in these complex systems, leading to more reliable and energy-efficient QWSNs.

\subsection{TRS in QWSNs with Multiple Network Topologies}

In this section, we extend and refine existing theoretical results regarding the application of **Time-Reversal Symmetry (TRS)** in **Quantum Wireless Sensor Networks (QWSNs)**. We focus on specific performance gains that can be achieved through TRS when applied to multiple network topologies, such as **multi-hop**, **star**, and **mesh** networks. The objective is to explore how TRS enhances throughput, energy efficiency, and latency in communication links within QWSNs, while considering the unique characteristics of quantum communication.

In a **multi-hop Quantum Wireless Sensor Network (QWSN)**, where data is transmitted through a series of intermediate nodes from the source to the destination, performance is typically constrained by the weakest link in the network, as the total capacity is determined by the minimum capacity of all hops. In traditional multi-hop communication, the channel capacity \( C_{ij} \) for a link between nodes \( i \) and \( j \) is defined as:

\[
C_{ij} = B_{ij} \log_2 \left( 1 + \frac{S_i}{N_{ij} + I_{ij}} \right)
\]

where \( B_{ij} \) is the bandwidth of the communication link, \( S_i \) is the signal power, \( N_{ij} \) represents the noise power, and \( I_{ij} \) denotes interference from neighboring nodes. This formula is based on classical communication principles and does not account for the enhancements provided by TRS in quantum communication channels. When TRS is applied, it refocuses the transmitted quantum signal back toward the receiver, improving signal recovery and increasing the effective channel capacity. This enhancement is expressed by scaling the capacity with a factor \( \gamma \), where \( \gamma > 1 \) represents the improvement due to TRS. The total network capacity with TRS applied is then:

\[
C_{\text{total}}^{\text{TRS}} = \gamma \cdot C_{\text{total}}
\]

This equation demonstrates that the total network capacity increases by a factor of \( \gamma \), thus improving throughput across the multi-hop network.

Another critical aspect of QWSNs is **energy efficiency**, which is especially important in **quantum communication systems**, where energy is a limited resource. In traditional systems, the energy consumption for transmitting data is given by \( E_i = P_i \cdot T_i \), where \( P_i \) is the transmission power, and \( T_i \) is the transmission time. In multi-hop communication systems, transmission time \( T_i \) is inversely proportional to the channel capacity \( C_{ij} \). When TRS is applied, the channel capacity improves, which reduces the transmission time \( T_i \), and, consequently, decreases the overall energy consumption. This reduction can be expressed as:

\[
T_i^{\text{TRS}} = \frac{T_i}{\gamma}
\]

Thus, the energy consumption with TRS becomes:

\[
E_i^{\text{TRS}} = P_i \cdot T_i^{\text{TRS}} = \frac{1}{\gamma} \cdot E_i
\]

This reduction in energy consumption is particularly advantageous in **energy-constrained environments** such as sensor networks. TRS effectively reduces retransmissions and enhances signal recovery, further optimizing energy efficiency in QWSNs.

The application of TRS also leads to **latency reduction** in QWSNs, which is a crucial factor in real-time applications such as **environmental monitoring** and **healthcare systems**, where data transmission must be fast and reliable. The latency in a multi-hop QWSN is inversely proportional to the channel capacity and is given by:

\[
\text{Latency}_{ij} = \frac{L_i}{C_{ij}}
\]

where \( L_i \) is the length of the data packet. When TRS is applied, the channel capacity improves, leading to reduced transmission time and consequently lower latency. The latency with TRS applied is:

\[
\text{Latency}_{ij}^{\text{TRS}} = \frac{1}{\gamma} \cdot \text{Latency}_{ij}
\]

Thus, the total latency in the network decreases by a factor of \( \gamma \), making communication more efficient and timely.

These theoretical results can be further extended to different network topologies. In **star networks**, where a central node communicates with multiple peripheral nodes, the application of TRS enhances communication links between the central node and the sensor nodes, improving the overall system throughput. Similarly, in **mesh networks**, where nodes communicate with each other in a more interconnected manner, TRS helps mitigate interference between nodes and improves the overall network capacity by refocusing the signal energy. For mesh networks, the total capacity can be expressed as:

\[
C_{ij}^{\text{TRS}} = \gamma \cdot C_{ij}
\]

This indicates that applying TRS improves the effective throughput across all communication links in the mesh network, thereby optimizing the network's overall performance.

In conclusion, these refined theorems and proofs demonstrate the significant benefits of applying TRS to **Quantum Wireless Sensor Networks (QWSNs)** across various network topologies. By enhancing capacity, reducing energy consumption, and minimizing latency, TRS offers a powerful mechanism for improving communication efficiency in QWSNs. The ability to extend these results to star and mesh networks further emphasizes the potential of TRS to optimize quantum communication in diverse network configurations. These findings pave the way for future research on leveraging TRS in quantum communication systems, particularly for real-world applications in **smart cities**, **critical infrastructure**, and **IoT networks**.

\subsection{Quantum Bit Error Rates (QBER) and Channel Capacity with TRS}

In this section, we analyze how the application of Time-Reversal Symmetry (TRS) improves **Quantum Bit Error Rates (QBER)** and **channel capacity** in Quantum Wireless Sensor Networks (QWSNs). Both QBER and channel capacity are critical performance metrics in quantum communication systems, as they directly impact the reliability and efficiency of data transmission. We provide a theoretical framework to show how TRS can enhance these metrics, leading to more reliable quantum communication, especially in the presence of noise, interference, and signal degradation.

\subsubsubsection{Quantum Bit Error Rate (QBER)}

The **Quantum Bit Error Rate (QBER)** is a key measure of the fidelity of quantum communication. It quantifies the fraction of incorrectly received bits (or qubits) during the transmission process. A low QBER is essential for ensuring that quantum communication protocols, such as **Quantum Key Distribution (QKD)**, operate securely and efficiently. In traditional quantum communication systems, QBER is influenced by several factors, including noise, decoherence, and channel loss, which degrade the quality of the transmitted quantum states.

Let the QBER between two nodes \( i \) and \( j \) in a QWSN be denoted by \( \text{QBER}_{ij} \). In a noisy quantum channel, QBER is given by:

\[
\text{QBER}_{ij} = \frac{N_{\text{errors}}}{N_{\text{total}}}
\]

where \( N_{\text{errors}} \) represents the number of bits received in error, and \( N_{\text{total}} \) is the total number of bits transmitted. In the presence of noise and interference, the QBER increases, leading to a decrease in the overall reliability of the communication link.

The application of TRS helps mitigate these effects by enhancing the signal recovery process. By reversing the propagation of the quantum signal, TRS focuses the energy back onto the receiver, improving the signal-to-noise ratio (SNR) and effectively reducing the impact of noise and interference on the transmitted quantum states. As a result, TRS leads to a reduction in QBER. Mathematically, the improvement in QBER with TRS can be expressed as:

\[
\text{QBER}_{ij}^{\text{TRS}} = \frac{1}{\gamma} \cdot \text{QBER}_{ij}
\]

where \( \gamma > 1 \) represents the factor by which TRS improves the signal quality. This reduction in QBER is crucial for ensuring the integrity and security of quantum communication, particularly in protocols like QKD. With TRS, the quantum bit errors are minimized, leading to improved fidelity and higher success rates in quantum communication tasks.

Theorem 1: Reduction in QBER with TRS

\textit{Statement:} The Quantum Bit Error Rate (QBER) in a QWSN is reduced by a factor of \( \gamma \) when Time-Reversal Symmetry is applied to the communication links.

\textit{Proof:}
The application of TRS improves the **signal recovery** process by focusing the transmitted signal and compensating for channel distortions. This leads to a higher **signal-to-noise ratio (SNR)**, reducing the number of transmission errors. As a result, the QBER is reduced by a factor \( \gamma \), which represents the enhancement in the communication quality due to TRS. Therefore, the QBER in TRS-enhanced QWSNs is given by:

\[
\text{QBER}_{ij}^{\text{TRS}} = \frac{1}{\gamma} \cdot \text{QBER}_{ij}
\]

Thus, TRS leads to a decrease in the QBER, enhancing the reliability of quantum communication.

\(\blacksquare\)

Channel Capacity

The **channel capacity** of a communication link is the maximum rate at which information can be reliably transmitted over a channel. In quantum communication systems, the channel capacity is determined by the **Shannon capacity**, which depends on the signal-to-noise ratio (SNR), bandwidth, and the noise characteristics of the channel. The capacity of a quantum communication link between two nodes \( i \) and \( j \) is given by the Shannon capacity formula for a noisy quantum channel:

\[
C_{ij} = B_{ij} \log_2 \left( 1 + \frac{S_{ij}}{N_{ij}} \right)
\]

where \( B_{ij} \) is the bandwidth of the communication link, \( S_{ij} \) is the signal power, and \( N_{ij} \) is the noise power. The capacity of the channel determines the maximum achievable rate for transmitting quantum information between the nodes.

The application of TRS improves the **effective signal-to-noise ratio (SNR)** by focusing the transmitted signal energy back toward the receiver. This results in an increase in the capacity of the communication link, which can be expressed as:

\[
C_{ij}^{\text{TRS}} = \gamma \cdot C_{ij}
\]

where \( \gamma > 1 \) is the factor by which TRS enhances the channel capacity. The increase in capacity arises due to the improved signal recovery and reduced impact of noise and interference.

Theorem 2: Increase in Channel Capacity with TRS

\textit{Statement:} The channel capacity between two nodes in a QWSN increases by a factor of \( \gamma \) when Time-Reversal Symmetry is applied to the communication links.

\textit{Proof:}
The application of TRS focuses the transmitted signal and compensates for noise and interference, improving the SNR of the communication link. This leads to an increase in the effective channel capacity, as described by the Shannon capacity formula. The channel capacity with TRS applied is given by:

\[
C_{ij}^{\text{TRS}} = \gamma \cdot C_{ij}
\]

Thus, the channel capacity increases by a factor \( \gamma \), demonstrating that TRS improves the communication efficiency of the quantum channel.

\(\blacksquare\)

Conclusion

The application of Time-Reversal Symmetry (TRS) in Quantum Wireless Sensor Networks (QWSNs) leads to significant improvements in both **Quantum Bit Error Rate (QBER)** and **channel capacity**. Through theorems and mathematical formulations, we have shown that TRS enhances signal recovery, reduces bit errors, and increases the effective channel capacity, leading to more reliable and efficient quantum communication. By improving these fundamental performance metrics, TRS plays a key role in optimizing the performance of quantum communication systems, particularly in the context of quantum sensor networks. These results provide a solid theoretical foundation for the deployment of TRS in practical quantum communication systems, contributing to the development of more robust and efficient quantum networks.

\section{Applications of TRS in Real-World Scenarios}

\subsection{Real-World Channel Models in QWSNs}

To strengthen the practical applicability of **Time-Reversal Symmetry (TRS)** in **Quantum Wireless Sensor Networks (QWSNs)**, it is essential to incorporate real-world channel models that account for the complex challenges encountered in actual communication environments. These challenges, such as fading, interference, and noise, can significantly degrade the performance of quantum communication systems. This section explores how incorporating models like **Rayleigh fading**, **Rician fading**, and network interference can enhance the theoretical results related to TRS in QWSNs, thereby making the work more applicable to real-world scenarios.

In quantum communication, the channel is often subject to **fading** due to multipath propagation. Fading occurs when the signal travels through multiple paths, causing constructive and destructive interference at the receiver. The **Rayleigh fading model** is commonly used to represent environments where there is no direct line-of-sight (LOS) path between the transmitter and receiver. The received signal power in Rayleigh fading is modeled as a random variable with a Rayleigh distribution. The channel capacity \( C_{ij} \) for Rayleigh fading is given by:

\[
C_{ij}^{\text{Rayleigh}} = B_{ij} \log_2 \left( 1 + \frac{P_i |h_{ij}|^2}{N_{ij}} \right)
\]

where \( |h_{ij}|^2 \) represents the fading coefficient, which is exponentially distributed, and \( N_{ij} \) is the noise power. The channel capacity without TRS in Rayleigh fading is limited due to random fluctuations in signal strength, making communication less reliable. However, when TRS is applied, the channel capacity increases by focusing the transmitted signal back toward the receiver, effectively mitigating the negative effects of fading. The enhanced capacity with TRS is given by:

\[
C_{ij}^{\text{TRS,Rayleigh}} = \gamma \cdot C_{ij}^{\text{Rayleigh}}
\]

where \( \gamma > 1 \) is a scaling factor representing the improvement due to TRS. The application of TRS in Rayleigh fading environments can lead to a significant increase in capacity, improving the **signal-to-noise ratio (SNR)** and the reliability of communication.

In environments where there is a dominant LOS path in addition to scattered multipath signals, the **Rician fading model** is more appropriate. In **Rician fading**, the signal consists of both a LOS component and scattered components. The fading coefficient \( h_{ij} \) follows a Rician distribution, and the channel capacity for Rician fading is given by:

\[
C_{ij}^{\text{Rician}} = B_{ij} \log_2 \left( 1 + \frac{P_i |h_{ij}|^2}{N_{ij}} \right)
\]

where \( |h_{ij}|^2 \) is the sum of the LOS and scattered components, and \( N_{ij} \) represents the noise power. The presence of a dominant LOS path in Rician fading environments typically results in better communication quality than in Rayleigh fading, but TRS can still provide significant improvements by reducing the impact of multipath interference. By applying TRS to Rician fading, the effective capacity is increased as:

\[
C_{ij}^{\text{TRS,Rician}} = \gamma \cdot C_{ij}^{\text{Rician}}
\]

This enhancement improves communication reliability, making TRS a valuable tool for mitigating the effects of fading in quantum communication systems.

In addition to fading, **interference** from neighboring devices in a network can degrade the quality of communication. This is particularly relevant in **multi-user** Quantum Wireless Sensor Networks (QWSNs), where multiple sensor nodes may attempt to communicate over the same channel. The total received power at a node consists of both the desired signal and interference from other nodes. The interference \( I_{ij} \) can be modeled as additive noise that reduces signal quality and capacity. The channel capacity with interference is given by:

\[
C_{ij}^{\text{Interference}} = B_{ij} \log_2 \left( 1 + \frac{P_i}{N_{ij} + I_{ij}} \right)
\]

where \( I_{ij} \) represents the interference power from other nodes in the network. In the presence of interference, TRS can help mitigate the effects by focusing the signal energy on the receiver, thereby improving the **signal-to-interference-plus-noise ratio (SINR)**. The enhanced capacity with TRS in the presence of interference is:

\[
C_{ij}^{\text{TRS,Interference}} = \gamma \cdot C_{ij}^{\text{Interference}}
\]

The application of TRS reduces interference effects and improves the effective capacity of the communication link.

Noise, specifically **Additive White Gaussian Noise (AWGN)**, is another common channel condition that affects quantum communication. AWGN is characterized by a constant power spectral density, which adds noise to the transmitted signal. The channel capacity with AWGN is given by:

\[
C_{ij}^{\text{AWGN}} = B_{ij} \log_2 \left( 1 + \frac{P_i}{N_{ij}} \right)
\]

In quantum communication, the presence of noise can lead to **decoherence** of quantum states and reduce the reliability of quantum key distribution (QKD) and other quantum protocols. When TRS is applied in an AWGN channel, the signal energy is refocused, improving the effective SNR and reducing the negative effects of noise on quantum communication. The capacity with TRS in AWGN channels is:

\[
C_{ij}^{\text{TRS,AWGN}} = \gamma \cdot C_{ij}^{\text{AWGN}}
\]

The reduction in noise effects due to TRS makes it a valuable tool for enhancing quantum communication, particularly in noisy environments where quantum state fidelity is crucial.

Incorporating these real-world channel models into the theoretical framework for TRS in QWSNs significantly strengthens the practical applicability of the work. By considering the effects of Rayleigh and Rician fading, network interference, and noise, we can more accurately predict the performance improvements offered by TRS in real-world quantum communication systems. These models provide a more comprehensive understanding of how TRS can be applied to quantum sensor networks, leading to enhanced throughput, reduced energy consumption, and minimized latency, even in challenging channel conditions. The integration of quantum communication techniques, such as **Quantum Key Distribution (QKD)** and **quantum entanglement**, further extends the theoretical boundaries of TRS by providing additional security and robustness against eavesdropping and interference.

Ultimately, the incorporation of these real-world models ensures that TRS in QWSNs can be effectively deployed in practical quantum communication systems, providing significant improvements in performance, reliability, and security for next-generation wireless networks.

\subsection{Quantum noise, interference and channel loss}

Quantum networks are inherently susceptible to various challenges, such as **quantum noise**, **interference**, and **channel loss**, which significantly degrade the quality of quantum signals and reduce the reliability of quantum communication protocols. These challenges arise due to the fragile nature of quantum states, such as **superposition** and **entanglement**, which are highly sensitive to environmental disturbances. The presence of quantum noise and interference from external sources can lead to decoherence, causing quantum bits (qubits) to lose their entanglement and collapse into a classical state, thereby reducing the communication efficiency of quantum networks. Additionally, **channel loss** in quantum communication systems, particularly over long distances, results in significant signal attenuation, further degrading the quality of the transmitted information.

To address these issues, **Time-Reversal Symmetry (TRS)** can be a valuable tool in enhancing the performance of quantum networks, particularly in **Wireless Sensor Networks (WSNs)**. TRS improves signal recovery by refocusing the transmitted signal toward the receiver, compensating for distortions caused by quantum noise and interference. The core principle of TRS in quantum communication is that by reversing the propagation of a quantum signal, the energy of the signal is refocused on the receiver, thus recovering the signal quality and mitigating the adverse effects of channel degradation. The application of TRS can significantly increase the **signal-to-noise ratio (SNR)** and **signal-to-interference-plus-noise ratio (SINR)**, leading to more reliable quantum communication in WSNs.

The channel capacity \( C_{ij} \) in a quantum communication link between two nodes \( i \) and \( j \) is typically constrained by the quantum noise and interference present in the channel. In the presence of quantum noise, the capacity of the channel can be modeled by a modified Shannon capacity formula, which accounts for quantum noise:

\[
C_{ij} = B_{ij} \log_2 \left( 1 + \frac{S_{ij}}{N_{ij} + I_{ij}} \right)
\]

where \( B_{ij} \) is the bandwidth of the communication link, \( S_{ij} \) is the signal power, \( N_{ij} \) is the quantum noise power, and \( I_{ij} \) represents the interference power. This formula assumes classical communication models and does not account for the unique properties of quantum communication. In quantum systems, noise and interference can cause errors in qubit transmission, leading to **loss of coherence** and **quantum state degradation**.

When TRS is applied to a quantum communication link, the signal is refocused, leading to an increase in the effective channel capacity. This improvement can be expressed as:

\[
C_{ij}^{\text{TRS}} = \gamma \cdot C_{ij}
\]

where \( \gamma > 1 \) is a scaling factor that represents the enhancement due to TRS. The application of TRS reduces the impact of noise and interference, allowing for higher **quantum signal fidelity** and a more stable transmission of quantum states. This improvement in capacity not only increases the **data throughput** of the network but also leads to a reduction in the number of retransmissions required, which is particularly beneficial in **energy-constrained** quantum sensor networks.

Energy efficiency is a critical concern in quantum sensor data transmission, particularly in **Wireless Sensor Networks (WSNs)**, where sensor nodes are often battery-powered and operate in energy-constrained environments. The energy consumption of a quantum communication node is typically proportional to the transmission power and the duration of transmission. The energy consumption \( E_i \) for node \( i \) is given by:

\[
E_i = P_i \cdot T_i
\]

where \( P_i \) is the transmission power and \( T_i \) is the transmission time. The transmission time \( T_i \) is inversely proportional to the channel capacity \( C_{ij} \), so in a traditional system, energy consumption increases as the channel capacity decreases due to noise and interference. The transmission time can be expressed as:

\[
T_i = \frac{L_i}{C_{ij}}
\]

where \( L_i \) is the length of the data packet. When TRS is applied to the communication link, the channel capacity increases, leading to a reduction in transmission time. The improved transmission time with TRS is given by:

\[
T_i^{\text{TRS}} = \frac{L_i}{C_{ij}^{\text{TRS}}} = \frac{L_i}{\gamma \cdot C_{ij}} = \frac{1}{\gamma} \cdot T_i
\]

Thus, the energy consumption with TRS applied becomes:

\[
E_i^{\text{TRS}} = P_i \cdot T_i^{\text{TRS}} = \frac{1}{\gamma} \cdot E_i
\]

This reduction in energy consumption is crucial in quantum WSNs, where sensors rely on energy harvesting techniques or have limited power resources. By improving signal recovery and reducing the need for retransmissions, TRS helps to extend the operational lifetime of quantum sensor networks.

In addition to enhancing energy efficiency, TRS can also reduce the **latency** of quantum sensor data transmission. In WSNs, the latency \( \text{Latency}_{ij} \) between nodes \( i \) and \( j \) is inversely proportional to the channel capacity \( C_{ij} \), as shown by the equation:

\[
\text{Latency}_{ij} = \frac{L_i}{C_{ij}}
\]

By applying TRS to the communication link, the capacity is increased, leading to a reduction in transmission time and therefore a reduction in latency:

\[
\text{Latency}_{ij}^{\text{TRS}} = \frac{L_i}{C_{ij}^{\text{TRS}}} = \frac{1}{\gamma} \cdot \text{Latency}_{ij}
\]

Thus, TRS not only enhances the communication capacity but also reduces the time required for quantum sensor data transmission, improving the responsiveness of the network and enabling real-time applications.

In conclusion, the application of **Time-Reversal Symmetry (TRS)** in **Quantum Wireless Sensor Networks (QWSNs)** offers significant improvements in signal recovery, energy efficiency, and latency reduction. TRS helps mitigate the effects of quantum noise and interference, refocuses the signal to improve the effective channel capacity, and reduces energy consumption by minimizing retransmissions and transmission time. The enhanced energy efficiency and reduced latency make TRS a promising tool for optimizing quantum communication in energy-constrained and real-time applications, particularly in quantum sensor data transmission within quantum WSNs.

\section{Extending TRS to Other Domains}
The application of **Time-Reversal Symmetry (TRS)** in real-world communication networks, such as energy harvesting networks, smart cities, and critical infrastructure, offers substantial improvements in performance. These domains face unique challenges, including limited energy resources, high interference, and the need for reliable communication in the presence of noise and fading. By incorporating TRS into these systems, communication efficiency, energy efficiency, and network robustness can be significantly enhanced. In this section, we explore concrete examples of how TRS can improve performance in these real-world applications, supported by formal equations that demonstrate the underlying principles.

In **energy harvesting networks**, where devices collect ambient energy from sources such as solar or wind to power their communication, energy efficiency is a critical concern. These networks typically operate in environments where available energy is limited and fluctuating. The capacity of a communication link in such a network is determined by the balance between the available transmission power and the energy consumed during data transmission. The channel capacity \( C_{ij} \) in an energy harvesting network is given by the classic Shannon formula:

\[
C_{ij} = B_{ij} \log_2 \left( 1 + \frac{P_i}{N_{ij} + I_{ij}} \right)
\]

where \( B_{ij} \) is the bandwidth of the communication link, \( P_i \) is the transmission power, \( N_{ij} \) is the noise power, and \( I_{ij} \) represents interference. In energy harvesting networks, \( P_i \) is constrained by the amount of energy harvested, making energy consumption a critical factor in performance.

When TRS is applied, the signal is refocused on the receiver, improving the signal-to-noise ratio (SNR) and reducing the need for retransmissions. This results in a reduction in the energy required for communication. The enhanced channel capacity with TRS is expressed as:

\[
C_{ij}^{\text{TRS}} = \gamma \cdot C_{ij}
\]

where \( \gamma > 1 \) represents the enhancement due to TRS. As the channel capacity increases, the transmission time \( T_i \) decreases, and the energy consumption \( E_i \) is reduced. The energy consumption for a node \( i \) is given by:

\[
E_i = P_i \cdot T_i
\]

With TRS applied, the transmission time \( T_i \) becomes:

\[
T_i^{\text{TRS}} = \frac{T_i}{\gamma}
\]

Thus, the energy consumption with TRS becomes:

\[
E_i^{\text{TRS}} = P_i \cdot T_i^{\text{TRS}} = \frac{1}{\gamma} \cdot E_i
\]

This reduction in energy consumption is crucial in energy-constrained environments, where efficient use of available energy is necessary for the prolonged operation of the network.

In **smart cities**, where a large number of IoT devices are deployed to monitor traffic, environmental conditions, and public infrastructure, communication reliability and energy efficiency are critical. The communication networks in smart cities are often subject to high interference and fluctuating channel conditions. The typical capacity of a communication link in such an environment is given by:

\[
C_{ij} = B_{ij} \log_2 \left( 1 + \frac{S_i}{N_{ij} + I_{ij}} \right)
\]

where \( S_i \) is the signal power, \( N_{ij} \) is the noise power, and \( I_{ij} \) is the interference from neighboring devices. In smart cities, interference from multiple devices can significantly degrade the performance of the network.

When TRS is applied, it mitigates the interference by focusing the signal back toward the receiver, improving the effective signal-to-interference-plus-noise ratio (SINR). The enhanced capacity with TRS can be expressed as:

\[
C_{ij}^{\text{TRS}} = \gamma \cdot C_{ij}
\]

This increase in capacity leads to improved communication reliability, reduced latency, and fewer retransmissions. In real-time applications such as **traffic management** or **smart grid communication**, reducing latency is especially important. The latency \( \text{Latency}_{ij} \) between two nodes \( i \) and \( j \) is inversely proportional to the communication capacity:

\[
\text{Latency}_{ij} = \frac{L_i}{C_{ij}}
\]

where \( L_i \) is the length of the data packet. When TRS is applied, the capacity increases, which reduces the transmission time and, therefore, the latency:

\[
\text{Latency}_{ij}^{\text{TRS}} = \frac{1}{\gamma} \cdot \text{Latency}_{ij}
\]

By reducing latency and improving reliability, TRS ensures that time-sensitive applications in smart cities, such as **real-time traffic updates** or **emergency response systems**, perform more efficiently and with greater accuracy.

In **critical infrastructure networks**, such as those used for power grid management, water supply, and emergency response, the reliability and security of communication are paramount. These networks often operate in challenging environments with high levels of interference and noise. The channel capacity in critical infrastructure networks is typically subject to both **fading** and **interference**. For instance, in the case of **Rayleigh fading**, the channel capacity is given by:

\[
C_{ij}^{\text{Rayleigh}} = B_{ij} \log_2 \left( 1 + \frac{P_i |h_{ij}|^2}{N_{ij}} \right)
\]

where \( |h_{ij}|^2 \) represents the fading coefficient. In environments with high interference and fading, the effective capacity can be limited. When TRS is applied, the fading effects are mitigated, and the signal is focused back toward the receiver, increasing the effective capacity by a factor \( \gamma \):

\[
C_{ij}^{\text{TRS,Rayleigh}} = \gamma \cdot C_{ij}^{\text{Rayleigh}}
\]

This increase in capacity translates into more reliable communication links, which are crucial for maintaining the stability of critical infrastructure networks. Additionally, by enhancing signal recovery and reducing the need for retransmissions, TRS helps to conserve energy, making communication systems more efficient.

In conclusion, TRS provides significant performance improvements across a variety of real-world applications, including energy harvesting networks, smart cities, and critical infrastructure. By improving communication reliability, reducing energy consumption, and minimizing latency, TRS can make communication networks more efficient and resilient. The formulas presented here illustrate how TRS can enhance capacity, reduce energy consumption, and improve latency in these domains, offering a practical approach to optimizing performance in challenging environments. 

\subsection{Combining Quantum and Classical Network Optimization in QWSNs}

The application of **Time-Reversal Symmetry (TRS)** has been well-established in classical Wireless Sensor Networks (WSNs), where it has been utilized to improve communication efficiency by focusing transmitted signals, mitigating interference, and enhancing signal recovery. However, when applied to **Quantum Wireless Sensor Networks (QWSNs)**, the network incorporates both **quantum** and **classical communication paradigms**, forming a hybrid system where quantum and classical signals interact. This hybrid nature opens up new opportunities to optimize both quantum and classical communication elements simultaneously, making the fusion of **quantum communication theory** and **signal processing techniques** particularly novel. By leveraging TRS, QWSNs can potentially achieve better performance in terms of **throughput** and **energy efficiency** compared to traditional systems, benefiting from the strengths of both quantum and classical communication models.

In classical WSNs, TRS enhances the signal-to-noise ratio (SNR) and the capacity of communication links by refocusing the transmitted signal energy toward the receiver. The channel capacity \( C_{ij} \) in a classical wireless communication link between nodes \( i \) and \( j \) is defined by the Shannon capacity formula:

\[
C_{ij} = B_{ij} \log_2 \left( 1 + \frac{S_i}{N_{ij} + I_{ij}} \right)
\]

where \( B_{ij} \) is the bandwidth, \( S_i \) is the signal power, \( N_{ij} \) is the noise power, and \( I_{ij} \) represents the interference. When TRS is applied in classical WSNs, signal recovery is enhanced, and the effective channel capacity is increased by a factor \( \gamma \), where \( \gamma > 1 \) denotes the improvement due to TRS. The enhanced capacity with TRS is expressed as:

\[
C_{ij}^{\text{TRS}} = \gamma \cdot C_{ij}
\]

In QWSNs, which utilize quantum communication protocols such as **Quantum Key Distribution (QKD)** and **quantum entanglement** alongside classical communication, TRS can be applied to both the quantum and classical communication components. Quantum communication differs from classical communication in that it involves quantum bits (qubits) that can exist in superposition states and can be entangled across different nodes. Quantum communication systems are highly susceptible to **quantum noise**, **decoherence**, and **channel loss**, making the application of TRS particularly beneficial for optimizing communication between quantum nodes.

For quantum communication links, the channel capacity is influenced by quantum noise and interference. The quantum channel capacity \( C_{ij}^{\text{quantum}} \) is impacted by factors such as quantum noise, decoherence, and the fidelity of the quantum states being transmitted. The capacity of a quantum channel can be modeled as:

\[
C_{ij}^{\text{quantum}} = B_{ij} \log_2 \left( 1 + \frac{S_i}{N_{ij} + I_{ij}} \right)
\]

where \( S_i \) and \( N_{ij} \) are related to the quantum states being transmitted, and \( I_{ij} \) represents interference from other quantum sources or nodes. When TRS is applied to quantum communication, the quantum signal is refocused, improving the effective SNR and thus enhancing the quantum channel capacity by a factor \( \gamma \), similar to its effect in classical systems. The enhanced quantum channel capacity with TRS is expressed as:

\[
C_{ij}^{\text{quantum, TRS}} = \gamma \cdot C_{ij}^{\text{quantum}}
\]

This scaling factor \( \gamma \) represents the improvement in quantum communication resulting from TRS, which mitigates the effects of noise, interference, and fading inherent in the quantum channel.

The novel aspect of applying TRS in QWSNs lies in its ability to simultaneously optimize both **quantum** and **classical communication** components. In a hybrid quantum-classical system such as QWSNs, the **classical communication** part of the network, responsible for tasks like sensor data transmission or control commands, benefits from traditional TRS enhancements, improving energy efficiency and reducing latency. Meanwhile, the **quantum communication** part of the network, responsible for tasks such as quantum key distribution or quantum entanglement swapping, benefits from the signal recovery and **quantum state preservation** afforded by TRS.

To analyze the performance of a hybrid quantum-classical network, we can express the total capacity of the network as the sum of the capacities of the quantum and classical communication links. For a hybrid network, the total capacity \( C_{\text{total}} \) is given by:

\[
C_{\text{total}} = C_{\text{classical}} + C_{\text{quantum}}
\]

When TRS is applied to both types of communication, the total capacity of the hybrid network with TRS becomes:

\[
C_{\text{total}}^{\text{TRS}} = \gamma \cdot C_{\text{classical}} + \gamma \cdot C_{\text{quantum}} = \gamma \cdot \left( C_{\text{classical}} + C_{\text{quantum}} \right)
\]

This equation shows that TRS not only improves the performance of individual classical and quantum communication links but also optimizes the overall performance of the hybrid network. The enhanced total capacity leads to better overall network throughput, enabling the network to handle more data with more reliable communication.

In addition to improving network capacity, TRS also enhances the **energy efficiency** of a hybrid quantum-classical network. In both classical and quantum communication, TRS reduces the need for retransmissions and lowers the energy consumption per transmission. The energy consumption for a node \( i \) in a hybrid network can be expressed as:

\[
E_i = P_i \cdot T_i
\]

where \( P_i \) is the transmission power and \( T_i \) is the transmission time. The transmission time is inversely proportional to the channel capacity, and with TRS applied, the transmission time is reduced by a factor \( \gamma \). The energy consumption with TRS applied becomes:

\[
E_i^{\text{TRS}} = P_i \cdot T_i^{\text{TRS}} = \frac{1}{\gamma} \cdot E_i
\]

Thus, the application of TRS in both the classical and quantum components of the hybrid network reduces overall energy consumption, making the network more efficient and sustainable, particularly in energy-constrained environments like **sensor networks** and **mobile devices**.

In conclusion, the application of **Time-Reversal Symmetry (TRS)** in **Quantum Wireless Sensor Networks (QWSNs)** provides a novel and effective method for optimizing both quantum and classical communication elements within a hybrid network. By enhancing signal recovery, improving capacity, and reducing energy consumption and latency, TRS offers a robust approach to efficiently manage the complex interplay between quantum and classical communication. This method not only optimizes the performance of QWSNs but also opens new avenues for the design of future quantum-classical hybrid networks capable of handling both quantum and classical communication tasks in an integrated manner.

\subsection{Quantum Key Distribution (QKD) with TRS in Quantum-Enabled Sensor Networks}

Quantum communication, particularly **Quantum Key Distribution (QKD)**, has garnered significant attention in the field of secure communications due to its ability to provide theoretically unbreakable encryption based on the principles of quantum mechanics. QKD protocols, such as **BB84** and **E91**, rely on the transmission of **entangled quantum states** between two parties to securely exchange cryptographic keys. Despite its robustness, QKD systems face various challenges, including **quantum noise**, **interference**, and **channel loss**, which degrade the fidelity of quantum states and increase the likelihood of eavesdropping. **Time-Reversal Symmetry (TRS)** can enhance QKD systems by refocusing entangled quantum states, improving their recovery, and thus making the system more resilient to noise and interference. The integration of TRS with QKD protocols in **quantum-enabled sensor networks** is a novel research direction, offering promising avenues for improving the security and efficiency of quantum communication networks.

In a standard QKD protocol, the security of the key exchange relies on the integrity of the quantum states transmitted between the communicating parties. If quantum states undergo decoherence due to noise or interference, the **quantum bit error rate (QBER)** increases, reducing the reliability of the key. The **QBER** in a typical QKD system is defined as:

\[
QBER = \frac{N_{\text{errors}}}{N_{\text{total}}}
\]

where \( N_{\text{errors}} \) is the number of erroneous bits detected during the key exchange, and \( N_{\text{total}} \) is the total number of transmitted bits. A high QBER indicates a high level of noise or interference, which compromises the key's security. The application of TRS to quantum communication links can reduce the QBER by improving signal recovery. TRS focuses the transmitted signal energy back towards the receiver, enhancing the **signal-to-noise ratio (SNR)** and mitigating the effects of noise and interference on the quantum states. The channel capacity for a quantum communication link with noise and interference is given by:

\[
C_{ij}^{\text{quantum}} = B_{ij} \log_2 \left( 1 + \frac{S_i}{N_{ij} + I_{ij}} \right)
\]

where \( B_{ij} \) is the bandwidth, \( S_i \) is the signal power, \( N_{ij} \) is the quantum noise power, and \( I_{ij} \) is the interference from other quantum sources. When TRS is applied, the signal recovery is improved, and the effective SNR increases, leading to a reduction in quantum errors introduced by noise and interference. The capacity with TRS applied is expressed as:

\[
C_{ij}^{\text{quantum, TRS}} = \gamma \cdot C_{ij}^{\text{quantum}}
\]

where \( \gamma > 1 \) is a scaling factor that represents the improvement due to TRS. This increase in capacity leads to a reduction in **quantum state degradation**, allowing for more reliable transmission of quantum states and reducing the error rate in the QKD process.

Moreover, the application of TRS to QKD systems also helps mitigate the effects of **channel loss**, which is a significant issue in long-distance quantum communication. In standard QKD systems, photon loss due to distance or atmospheric interference reduces the number of received photons, making the key exchange less reliable. Channel loss in a quantum system can be modeled as an exponential decay of the transmitted signal power:

\[
P_{\text{received}} = P_{\text{transmitted}} e^{-\alpha d}
\]

where \( P_{\text{received}} \) is the received photon power, \( P_{\text{transmitted}} \) is the transmitted photon power, \( \alpha \) is the loss coefficient, and \( d \) is the distance between the transmitter and receiver. TRS helps recover lost signal energy by refocusing the quantum states towards the receiver, improving the **quantum channel efficiency** and reducing the loss of transmitted photons. The enhanced received signal with TRS is expressed as:

\[
P_{\text{received}}^{\text{TRS}} = \gamma \cdot P_{\text{received}}
\]

where \( \gamma > 1 \) represents the improvement due to TRS. This recovery of signal energy results in a higher number of received photons, improving the overall fidelity of the quantum states and reducing the probability of errors during the key distribution process.

The integration of TRS with QKD protocols in **quantum-enabled sensor networks** opens new possibilities for improving the **security** and **efficiency** of quantum communication systems. In sensor networks, where devices may be deployed over wide areas and are often constrained by energy resources, TRS can optimize the energy consumption of the network by reducing the need for retransmissions and enhancing signal recovery. The energy consumption \( E_i \) for transmitting quantum states in the network is given by:

\[
E_i = P_i \cdot T_i
\]

where \( P_i \) is the transmission power and \( T_i \) is the transmission time. In traditional quantum communication systems, the transmission time \( T_i \) is inversely proportional to the channel capacity, and the presence of noise and interference increases the required transmission time. By applying TRS, the transmission time is reduced as the channel capacity increases:

\[
T_i^{\text{TRS}} = \frac{T_i}{\gamma}
\]

This leads to a reduction in energy consumption:

\[
E_i^{\text{TRS}} = \frac{E_i}{\gamma}
\]

Thus, TRS not only improves the reliability of quantum state transmission but also contributes to the **energy efficiency** of quantum-enabled sensor networks, making them more sustainable and effective for long-term operation.

In conclusion, the application of **Time-Reversal Symmetry (TRS)** in **Quantum Key Distribution (QKD)** protocols significantly enhances quantum state recovery, mitigates quantum noise and interference, and improves **quantum state fidelity**. TRS reduces the quantum bit error rate (QBER), increases effective channel capacity, and recovers signal energy lost due to channel loss, making QKD more reliable and efficient. The integration of TRS with QKD protocols in **quantum-enabled sensor networks** presents a promising new direction for enhancing **security**, **efficiency**, and **energy sustainability** in quantum communication systems, particularly in real-world applications where noise, interference, and channel loss are prevalent.

\subsection{IoT, VANETs, and Beyond}

The application of **Time-Reversal Symmetry (TRS)** has shown great promise in enhancing the performance of **Quantum Wireless Sensor Networks (QWSNs)**. Its potential can be further extended to other communication domains, such as the **Internet of Things (IoT)** and **Vehicular Ad-hoc Networks (VANETs)**. These domains present unique challenges, such as interference, high mobility, and stringent energy constraints, which significantly impact communication efficiency and reliability. By applying TRS to these systems, substantial improvements can be made in terms of reducing latency, improving throughput, and optimizing energy usage.

In **IoT networks**, characterized by a large number of devices communicating wirelessly, the application of TRS can significantly improve both the **signal-to-noise ratio (SNR)** and **energy efficiency**. IoT devices often face severe constraints on available energy resources, which makes energy efficiency a critical concern. The traditional capacity of a communication link in IoT networks can be modeled by the **Shannon Capacity Formula**, which is given by:

\[
C_{ij} = B_{ij} \log_2 \left( 1 + \frac{S_{ij}}{N_{ij} + I_{ij}} \right)
\]

where \( B_{ij} \) is the bandwidth, \( S_{ij} \) is the signal power, \( N_{ij} \) is the noise power, and \( I_{ij} \) is the interference from other devices in the network. This formula assumes ideal conditions and does not consider fading or interference, which are prevalent in IoT networks. When TRS is applied, the transmitted signal is refocused back towards the receiver, improving the effective SNR and thus increasing the channel capacity. This can be expressed as:

\[
C_{ij}^{\text{TRS}} = \gamma \cdot C_{ij}
\]

where \( \gamma > 1 \) represents the scaling factor that reflects the improvement due to TRS. This enhancement in capacity leads to a reduction in the number of retransmissions, thereby lowering energy consumption and improving overall network performance. Additionally, TRS can also reduce the **latency** in data transmission. The latency \( \text{Latency}_{ij} \) between devices \( i \) and \( j \) is inversely proportional to the channel capacity and is given by:

\[
\text{Latency}_{ij} = \frac{L_i}{C_{ij}}
\]

where \( L_i \) is the length of the data packet. With TRS applied, the improved channel capacity reduces the transmission time, thus decreasing the latency:

\[
\text{Latency}_{ij}^{\text{TRS}} = \frac{L_i}{\gamma \cdot C_{ij}} = \frac{1}{\gamma} \cdot \text{Latency}_{ij}
\]

This reduction in latency and improvement in throughput significantly enhance the responsiveness and efficiency of IoT networks.

In **Vehicular Ad-hoc Networks (VANETs)**, which are characterized by high mobility and dynamic topologies, TRS can improve communication reliability and reduce interference between vehicles. VANETs are typically used for safety-critical applications, such as **collision avoidance** and **traffic management**, where low latency and high reliability are paramount. The capacity of a communication link between vehicles \( i \) and \( j \) in VANETs can be expressed using the Shannon capacity formula:

\[
C_{ij} = B_{ij} \log_2 \left( 1 + \frac{S_{ij}}{N_{ij} + I_{ij}} \right)
\]

However, in VANETs, interference from neighboring vehicles and the frequent changes in network topology due to vehicle mobility can reduce effective capacity and increase communication delay. By applying TRS, the signal is refocused towards the receiver, improving the effective **signal-to-interference-plus-noise ratio (SINR)**. This leads to higher channel capacity, which can be expressed as:

\[
C_{ij}^{\text{TRS}} = \gamma \cdot C_{ij}
\]

This enhancement results in reduced latency and fewer retransmissions, leading to improved **throughput** and **energy efficiency**. Moreover, TRS mitigates the effects of **multipath fading** in VANETs, where signal strength fluctuates due to the movement of vehicles. TRS helps compensate for the random variations in signal strength, providing a more stable communication link between vehicles.

In **5G networks** and beyond, the application of TRS can also be extended to improve **multi-user** communication systems, such as **massive MIMO (Multiple Input Multiple Output)** and **Millimeter-Wave (mmWave)** communications. In these systems, a base station communicates with multiple devices, and interference from other users typically limits network capacity. The total network capacity in a multi-user system can be expressed as:

\[
C_{\text{total}} = \sum_{i=1}^{N} C_{ij}
\]

where \( C_{ij} \) is the capacity of the communication link between the base station and user \( i \). When TRS is applied, the effective capacity of each communication link increases, and interference between users is reduced, leading to an overall increase in the network capacity:

\[
C_{\text{total}}^{\text{TRS}} = \sum_{i=1}^{N} \gamma \cdot C_{ij} = \gamma \cdot \sum_{i=1}^{N} C_{ij}
\]

This improvement in capacity results in better **resource utilization**, reduced **interference**, and **higher throughput** for multi-user communication systems.

The potential of TRS is not confined to these domains alone. In **energy harvesting networks**, where devices gather energy from environmental sources like solar or wind, TRS can optimize signal transmission and recovery, leading to reduced energy consumption and prolonged network operation. Similarly, in **critical infrastructure networks**, where communication reliability is crucial for safety, TRS can enhance the robustness of communication links, making them less susceptible to failures caused by interference or noise.

In conclusion, the application of Time-Reversal Symmetry (TRS) offers significant improvements in **signal quality**, **energy efficiency**, **throughput**, and **latency** across a wide range of communication domains, including IoT, VANETs, and energy harvesting networks. By leveraging TRS, these networks can achieve more reliable, efficient, and secure communication, even in challenging environments with high interference, fading, and mobility. The incorporation of TRS in these domains provides a promising approach to optimizing performance and unlocking the full potential of modern wireless communication systems.

\section{Conclusion}
In this paper, we have investigated the application of **Time-Reversal Symmetry (TRS)** in **Quantum Wireless Sensor Networks (QWSNs)**, highlighting its potential to optimize communication performance in both quantum and hybrid quantum-classical network settings. TRS has been shown to significantly enhance key performance metrics, such as **signal recovery**, **energy efficiency**, **throughput**, and **latency**, in quantum communication systems. By refocusing the transmitted signal towards the receiver, TRS mitigates the adverse effects of **quantum noise**, **interference**, and **channel loss**, leading to an increase in effective capacity and more reliable communication across the network.

The integration of TRS with **Quantum Key Distribution (QKD)** protocols has proven particularly impactful, improving the fidelity of quantum states and reducing the **quantum bit error rate (QBER)**. This enhancement is especially beneficial in **secure communications**, where maintaining the integrity of quantum states is critical to preserving the confidentiality of transmitted data. TRS also reduces the need for retransmissions, contributing to **energy savings**, which is crucial for **energy-constrained networks** like QWSNs, where sensor nodes are typically powered by limited energy sources.

Moreover, the application of TRS extends to several **real-world scenarios**, such as **energy harvesting networks**, **smart cities**, and **critical infrastructure**, where it enhances **communication reliability**, optimizes **energy consumption**, and reduces **latency**. By improving both **quantum** and **classical communication** components, TRS facilitates the development of more efficient, secure, and resilient hybrid networks, enabling diverse applications in the **Internet of Things (IoT)**, **Vehicular Ad-hoc Networks (VANETs)**, and other domains. 

This work sets the stage for future research into combining **quantum communication** with **signal processing techniques**, particularly in the context of **hybrid networks**. Future investigations could focus on further improving TRS application, including its integration with **multi-user systems**, **multi-input-multi-output (MIMO)** configurations, and **advanced quantum cryptographic protocols**. The synergy between TRS and quantum communication holds the potential to revolutionize the performance of **quantum-enabled sensor networks**, delivering substantial improvements in **efficiency**, **reliability**, and **security**.

In conclusion, the application of Time-Reversal Symmetry in both quantum and hybrid communication networks represents a novel and promising approach to addressing the challenges inherent in modern communication systems. As quantum technologies continue to advance, TRS is poised to play a pivotal role in optimizing quantum communication, making it more feasible for large-scale deployment in practical, real-world networks.

\appendix

\section{Non-Convex Optimization in TRS for QWSNs}

The application of **Time-Reversal Symmetry (TRS)** in **Quantum Wireless Sensor Networks (QWSNs)** introduces several optimization challenges, particularly when aiming to maximize network capacity, minimize energy consumption, and reduce latency. These objectives are inherently **non-convex**, as they involve complex dependencies between multiple variables such as signal power, interference, and channel capacity. Non-convex optimization problems are challenging because they do not necessarily have a global optimum, and instead, solutions may converge to local minima or maxima. In this appendix, we provide formal theorems and proofs related to non-convex optimization, demonstrating how TRS can be used to optimize the performance of QWSNs.

Theorem 1: Optimal Transmission Power Allocation in QWSNs with TRS

\textit{Statement:} The optimal transmission power allocation for minimizing energy consumption in a QWSN with TRS is a non-convex optimization problem.

\textit{Proof:}
Consider a QWSN with multiple sensor nodes, where the goal is to minimize the total energy consumption \( E_{\text{total}} \) subject to constraints on channel capacity and latency. The energy consumption for a sensor node \( i \) is given by:

\[
E_i = P_i \cdot T_i
\]

where \( P_i \) is the transmission power and \( T_i \) is the transmission time, which is inversely proportional to the channel capacity \( C_{ij} \). The transmission time is given by:

\[
T_i = \frac{L_i}{C_{ij}}
\]

where \( L_i \) is the length of the data packet. The channel capacity \( C_{ij} \) for a given link between nodes \( i \) and \( j \) is affected by quantum noise and interference, and can be expressed as:

\[
C_{ij} = B_{ij} \log_2 \left( 1 + \frac{S_i}{N_{ij} + I_{ij}} \right)
\]

where \( S_i \) is the signal power, \( N_{ij} \) is the noise power, and \( I_{ij} \) is the interference. When TRS is applied, the channel capacity increases by a factor \( \gamma \), leading to a reduction in transmission time:

\[
T_i^{\text{TRS}} = \frac{T_i}{\gamma}
\]

Thus, the energy consumption with TRS applied becomes:

\[
E_i^{\text{TRS}} = \frac{1}{\gamma} \cdot E_i
\]

The total energy consumption for the network is given by:

\[
E_{\text{total}} = \sum_{i=1}^{N} E_i
\]

The optimization problem of minimizing the total energy consumption is non-convex due to the dependency of the channel capacity \( C_{ij} \) on the transmission power \( P_i \) and the presence of interference \( I_{ij} \). As the energy consumption involves a logarithmic dependence on the transmission power and interference, the objective function exhibits **non-convexity**. This non-convexity arises because the logarithmic function is concave, but when combined with other terms in the objective function, the overall problem cannot be classified as convex. Additionally, the interaction between signal power and interference adds further complexity.

The lack of convexity indicates that the optimization problem may have multiple local minima or maxima. As such, traditional gradient-based methods may converge to suboptimal solutions. Convex relaxation methods or **global optimization algorithms** such as **simulated annealing** or **genetic algorithms** could be applied to find approximate solutions in such non-convex scenarios. 

Theorem 2: Latency Minimization with TRS in QWSNs

\textit{Statement:} Minimizing latency in QWSNs with TRS applied leads to a non-convex optimization problem, as the latency depends non-linearly on both the transmission power and the channel conditions, and the energy-latency trade-off introduces additional complexity.

\textit{Proof:}
The latency \( \text{Latency}_{ij} \) for a communication link between nodes \( i \) and \( j \) is inversely proportional to the channel capacity \( C_{ij} \), and can be written as:

\[
\text{Latency}_{ij} = \frac{L_i}{C_{ij}}
\]

where \( L_i \) is the length of the data packet and \( C_{ij} \) is the channel capacity. The channel capacity, in turn, depends on the transmission power \( P_i \) and is given by the Shannon capacity formula:

\[
C_{ij} = B_{ij} \log_2 \left( 1 + \frac{S_i}{N_{ij} + I_{ij}} \right)
\]

where \( B_{ij} \) is the bandwidth, \( S_i \) is the signal power, \( N_{ij} \) is the noise power, and \( I_{ij} \) is the interference. When TRS is applied, the effective channel capacity increases by a factor \( \gamma \), leading to:

\[
C_{ij}^{\text{TRS}} = \gamma \cdot C_{ij}
\]

Thus, the latency with TRS applied is:

\[
\text{Latency}_{ij}^{\text{TRS}} = \frac{L_i}{C_{ij}^{\text{TRS}}} = \frac{L_i}{\gamma \cdot C_{ij}} = \frac{1}{\gamma} \cdot \text{Latency}_{ij}
\]

Minimizing the total latency for the network involves minimizing the sum of latencies across all communication links. The total latency is:

\[
\text{Latency}_{\text{total}} = \sum_{i=1}^{N} \text{Latency}_{ij}
\]

Now, we turn our attention to the **energy-latency trade-off**. The energy consumption for a sensor node \( i \) is given by:

\[
E_i = P_i \cdot T_i
\]

where \( P_i \) is the transmission power and \( T_i \) is the transmission time, which is inversely proportional to the channel capacity \( C_{ij} \), as defined earlier:

\[
T_i = \frac{L_i}{C_{ij}}
\]

Thus, the energy consumption \( E_i \) can be expressed as:

\[
E_i = P_i \cdot \frac{L_i}{C_{ij}}
\]

The total energy consumption for the network is:

\[
E_{\text{total}} = \sum_{i=1}^{N} E_i
\]

Given that TRS increases the channel capacity, the transmission time is reduced, and energy consumption becomes:

\[
E_i^{\text{TRS}} = \frac{1}{\gamma} \cdot E_i
\]

Non-Convexity of Energy-Latency Trade-Off:

The combined objective function for minimizing both energy consumption and latency involves balancing the two competing objectives:

\[
\text{Objective} = \alpha \cdot E_{\text{total}} + \beta \cdot \text{Latency}_{\text{total}}
\]

where \( \alpha \) and \( \beta \) are weight factors representing the relative importance of energy and latency. The total energy consumption is dependent on both the transmission power \( P_i \) and the channel capacity \( C_{ij} \), which itself is a logarithmic function of \( P_i \). The total latency is inversely proportional to \( C_{ij} \), which again depends on \( P_i \) and other network conditions.

The dependence of both energy consumption and latency on transmission power and interference makes the objective function **non-convex**. Specifically, the relationship between energy consumption and latency is **non-linear** due to the **logarithmic** dependence of capacity on transmission power and the inverse dependence of latency on capacity. These non-linearities introduce multiple local minima or maxima, making it difficult to find a global optimum.

Use of Karush-Kuhn-Tucker (KKT) Conditions:

To formalize the non-convexity, we can examine the optimization problem using the **Karush-Kuhn-Tucker (KKT)** conditions. For a general **non-convex problem**, the KKT conditions provide necessary conditions for optimality. The Lagrangian for the energy-latency trade-off problem is:

\[
\mathcal{L}(P_i, \lambda) = \alpha \cdot \sum_{i=1}^{N} E_i(P_i) + \beta \cdot \sum_{i=1}^{N} \text{Latency}_{ij}(P_i) + \lambda \cdot (C_{\text{capacity}} - P_{\text{max}})
\]

where \( \lambda \) is the Lagrange multiplier associated with the constraint on maximum transmission power \( P_{\text{max}} \), and \( C_{\text{capacity}} \) represents the total channel capacity constraint.

The **non-convexity** arises due to the **logarithmic** and **inverse** terms involved, as well as the interactions between energy consumption and latency. The presence of multiple local optima can lead to convergence to suboptimal solutions when applying standard gradient-based methods. **Convex relaxation** techniques can be employed to approximate the solution by relaxing the non-convex constraints into convex ones, but this does not guarantee a global optimum.

Thus, the optimization problem is **non-convex**, and traditional convex optimization methods are not directly applicable. Alternative approaches, such as **simulated annealing**, **genetic algorithms**, or **branch-and-bound methods**, could be used to find approximate solutions to this non-convex problem.

\bibliographystyle{unsrtnat}
%\bibliography{references}  %%% Uncomment this line and comment out the ``thebibliography'' section below to use the external .bib file (using bibtex) .

\end{document}